\documentclass[prd,aps]{revtex4}
\usepackage{amsmath}
\usepackage{amssymb}
\usepackage{amsfonts}
\usepackage{graphicx,bm}
\usepackage{dcolumn}
\usepackage[colorlinks=true]{hyperref}
\usepackage{epsf}
\usepackage{enumerate}
\usepackage{hhline}
\usepackage{array}
\usepackage{tabularx}
\usepackage{subfigure}

\newcommand{\be}{\begin{equation}}
\newcommand{\ee}{\end{equation}}
\newcommand{\bea}{\begin{eqnarray}}
\newcommand{\eea}{\end{eqnarray}}
\newcommand{\beaa}{\begin{eqnarray*}}
\newcommand{\eeaa}{\end{eqnarray*}}

\newcommand{\nn}{\nonumber \\}

\def\be{\begin{equation}}
\def\ee{\end{equation}}
\def\bea{\begin{eqnarray}}
\def\eea{\end{eqnarray}}

\begin{document}
\title{$3+1$ decomposition in modified gravities within the Palatini formalism and some applications}
\author{Diego S\'aez-Chill\'on G\'omez}
\email{diego.saez@uva.es} \affiliation{Department of Theoretical Physics, Atomic and Optics, Campus Miguel Delibes, \\ University of Valladolid UVA, Paseo Bel\'en, 7,
47011 - Valladolid, Spain}

\begin{abstract}

In the present paper, the $3+1$ decomposition of the spacetime onto hypersurface(s) is analysed and established for theories within the Palatini formalism by considering a general function of the Ricci scalar in the gravitational action. The corresponding Gauss-Codazzi relations are obtained and the boundary term that has to be subtracted in the gravitational action is easily deduced. Then, these relations are applied to the so-called ADM decomposition to describe the foliation of the spacetime onto hypersurfaces of constant time within these theories. Finally, the junction conditions are also obtained by using a decomposition in Gaussian normal coordinates, which coincide with the conditions deduced previously through different approaches.

\end{abstract}
%
%
\maketitle
%
%
%
\section{Introduction}
General Relativity (GR) and other gravitational theories are constructed in such way that the equivalence principle and specially general covariance are preserved. To do it so, gravity is described by spacetime geometry, which is completely defined through lengths and parallel transportation of vectors. The former is given by the spacetime metric while the latter is described by covariant derivatives, which are defined in terms of a connection. By constructing the appropriate scalar invariants that contain the metric, the connection and their derivatives, the gravitational action is given by the integral over the spacetime volume. Nevertheless, the corresponding covariant derivative is not unique but depends on the particular gauge prescription that is assumed, preserving in any case the invariance of the theory under general diffeomorphism. In this sense, it is well known the equivalence of GR as described by the Levi-Civita connection, which is torsionless and provides non-null curvature spacetimes, and the so-called Teleparallel version of GR that assumes the Weitzenb\"ock connection that leads to non-null torsion  but null curvature \cite{Aldrovandi:2013wha}.  In this sense, depending on the corresponding prescription, the gravitational action can be constructed in terms of scalar invariants departing from the Riemann tensor, the torsion tensor or the non-metricity tensor, all leading to the same dynamics when the gravitational action just contains linear functions of the above scalars \cite{BeltranJimenez:2017tkd}. In the so-called Palatini formalism, the shape of the connection is not assumed a priori but is considered as an independent field from the metric, with the most general affine connection containing a non-symmetric part (contortion) and non-metricity part (disformation). Nevertheless, by the projective invariance of the Ricci scalar, any gravitational action constructed in terms of the Ricci scalar will just depend on the torsionless part of the connection \cite{Olmo:2011uz,DeFelice:2010aj}, recovering automatically the Levi-Civita connection for the Hilbert-Einstein action. However, by considering a more general function of the Ricci scalar at the level of the action, the corresponding field equations do not lead to the Levi-Civita connection compatible with the spacetime metric but to a connection that is compatible to a conformal related metric \cite{Olmo:2011uz}. This is the natural extension of the so-called metric $f(R)$ gravities \cite{Nojiri:2017ncd} to the Palatini formalism, which have been widely considered in the literature in multiple frameworks (for a review see \cite{Olmo:2011uz,BeltranJimenez:2017doy,Vitagliano:2010sr}).   \\

Among the extensive analysis and applications of $f(\mathcal{R})$ gravities within the Palatini formalism, one might highlight the construction of cosmological solutions that circumvent the problem of dark energy in gravitational terms \cite{Baghram:2009we,Aoki:2018lwx,Leanizbarrutia:2017xyd,Rosa:2017jld,Rosa:2019ejh,Rosa:2021ish,Harko:2011nh} and the analysis of the inflationary paradigm within these theories \cite{Shimada:2018lnm,Gialamas:2020snr,Antoniadis:2018ywb,Enckell:2018hmo,Edery:2019txq,Rasanen:2018ihz,Bekov:2020dww} together with preheating/reheating \cite{Karam:2021sno,Gialamas:2019nly} and the growth of cosmological perturbations \cite{Koivisto:2005yc}. In addition, these modified gravities are inspired by Born-Infeld electromagnetism \cite{BeltranJimenez:2017doy}, as some regular black hole solutions can be obtained \cite{Olmo:2015axa}, also in combination with non-linear electrodynamics \cite{Guerrero:2020uhn}, Kerr black holes might be stable under perturbations in some hybrid versions of the theory \cite{Rosa:2020uoi}, and also some regular cosmologies can be constructed  \cite{Odintsov:2014yaa}. Some other frameworks that have been analysed within the Palatini formalism include models for the stellar structure \cite{Olmo:2019flu}, the analysis of the Cauchy problem \cite{Capozziello:2010ut} and the extension of the formalism to flat geometries \cite{BeltranJimenez:2018vdo}, among many others. \\

Here we aim to extend the well known $3+1$ decomposition of the spacetime to $f(\mathcal{R})$ gravities within the Palatini formalism and apply such decomposition to some frameworks. The $3+1$ decomposition consists on defining an hypersurface or a family of hypersurfaces that slices the spacetime and then express some geometrical variables as the Ricci tensor or the curvature, restricted to the hypersurface, in terms of the intrinsic and extrinsic curvatures of the hypersurface, as well as of the normal vector to the hypersurface(s) and its derivatives, leading to the so-called Gauss-Codazzi relations \cite{Gourgoulhon:2007ue,Lecturenotes}. Besides the proper geometrical aspect, such decomposition has multiple applications in gravitation, from the Arnowitt-Deser-Misner (ADM) decomposition of the spacetime onto hypersurfaces of constant time \cite{Arnowitt:1962hi}, which is the starting point to construct the hamiltonian formulation of GR, to the boundary terms that arise in the gravitational action or the junction conditions on a hypersurface that matches different spacetimes regions or branes. The framework has been previously explored in the scalar-tensor picture of $f(\mathcal{R})$ gravities in Ref.~\cite{Bombacigno:2019nua}. In this paper, we obtain explicitly the Gauss-Codazzi relations for Palatini $f(\mathcal{R})$ gravities, what leads to the Gauss-Codazzi action, where the corresponding boundary term, analog to the Gibbons-Hawking-York term in GR, is subtracted in the action, coinciding with the one obtained in \cite{Obukhov1987} for a general metric-affine theory for the Hilbert-Einstein action and the one found in \cite{Gomez:2020rnq} for $f(\mathcal{R})$ gravities by a well-posed variational principle, being also discussed in the framework of Brans-Dicke-like theories in \cite{Gionti:2020cwu}.  In addition, the ADM decomposition is also studied by applying the previous Gauss-Codazzi equations. Finally, the corresponding junction conditions for $f(\mathcal{R})$ gravities within the Palatini formalism are established, which coincide with the results obtained in \cite{Olmo:2020fri} through distributional analysis, and which have been shown to have important consequences in different scenarios, as the deflection of light by compact objects and the formation of (double) shadows \cite{Guerrero:2021pxt}.\\

The paper is organised as follows: in section \ref{backgroud} the main features of Palatini $f(\mathcal{R})$ gravity is reviewed. Section \ref{tresmasuno} is devoted to the $3+1$ decomposition in these theories and the corresponding Gauss-Codazzi relations. In section \ref{ADM}, the ADM decomposition is studied, while in Sect. \ref{Junctions} the corresponding junction conditions are obtained. Finally, the section \ref{Conclusions} summarises the results of the paper.

\section{Modified Palatini gravity}
\label{backgroud}
The Palatini approach consists on dealing with the spacetime metric and the connection as independent fields. For the Hilbert-Einstein action, such approach directly leads to the metricity condition, such that the connection reduces to the Levi-Civita one plus a projective mode and consequently GR field equations are recovered. Nevertheless, this is not the case for a more general function of the curvature scalar $\mathcal{R}$. Here we are considering the class of theories described by the following gravitational action:
\be
S=\frac{1}{2\kappa^2}\int dx^4 \sqrt{-g} \left[f(\mathcal{R}) +L_m\right] \ ,
\label{fRaction}
\ee
where $L_m$ is the matter Lagrangian which is assumed to depend solely on the metric and matter fields. The Ricci scalar $\mathcal{R}$ is defined by the contraction of the Ricci tensor with the spacetime metric $g_{\mu\nu}$:
\be
\mathcal{R}=g^{\mu\nu}\mathcal{R}_{\mu\nu}(\Gamma)\ ,
\ee
whereas the Ricci tensor is expressed in terms of the connection as:
\be
\mathcal{R}_{\mu\nu}(\Gamma)=\partial_{\lambda}\Gamma^{\lambda}_{\mu\nu}-\partial_{\nu}\Gamma^{\lambda}_{\mu\lambda}+\Gamma^{\lambda}_{\sigma\lambda}\Gamma^{\sigma}_{\mu\nu}-\Gamma^{\lambda}_{\sigma\nu}\Gamma^{\sigma}_{\mu\lambda}\ . 
\label{Ricci}
\ee
In the Palatini formalism, the connection $\Gamma$ is in principle independent of the spacetime metric. Nevertheless, one can assume that the connection that enters in the gravitational action through the Ricci tensor is torsionless because of the projective invariance of the Ricci scalar \cite{Gomez:2020rnq}. The field equations corresponding to variations of the action (\ref{fRaction}) with respect to the spacetime metric are \cite{Olmo:2011uz}:
\be
  f_{\mathcal{R}}\mathcal{R}_{\mu\nu}-\frac{1}{2}g_{\mu\nu}f=\kappa^2 T_{\mu\nu}\ .
  \label{Fieldeqs1}
  \ee
Here $T_{\mu\nu}=-\frac{2}{\sqrt{-g}}\frac{\delta  (\sqrt{-g} L_m)}{\delta g^{\mu\nu}}$ is the energy-momentum tensor and  $f_{\mathcal{R}}=\frac{df}{d\mathcal{R}}$. Variations of the gravitational action with respect to the connection $\Gamma$ provide the other set of field equations \cite{Olmo:2011uz}:
\be
  \tilde{\nabla}_{\lambda}\left(\sqrt{-g}f_{\mathcal{R}}g^{\mu\nu}\right)=0\ ,
  \label{Fieldeqs}
  \ee
where $\tilde{\nabla}$ is the covariant derivative defined by the connection $\Gamma$. The equation (\ref{Fieldeqs}) provides directly the expression for the connection $\Gamma$, as imposes to be compatible with a metric that is related to the spacetime metric by a conformal transformation as follows:
\be
q_{\mu\nu}=\Omega^2 g_{\mu\nu}\ , \quad \Omega^2=f_{\mathcal{R}}\ ,
\label{conform_transform22}
\ee
Hence, the field equations (\ref{Fieldeqs}) become:
\be
\tilde{\nabla}_{\lambda}\left(\sqrt{-q}q^{\mu\nu}\right)=0\ ,
\label{hmetricity}
\ee
which denotes the metricity condition of the covariant derivative $\tilde{\nabla}$ with respect to the metric $q_{\mu\nu}$.  Moreover, the trace of the field equations (\ref{Fieldeqs1}) establishes an algebraic relation among the scalar curvature $\mathcal{R}$ and the trace of the energy-momentum tensor $T$ that is given by:
\be
  f_{\mathcal{R}}\mathcal{R}-2f=\kappa^2 T\ .
   \label{traceRT}
   \ee
This is an algebraic equation that allows to obtain $\mathcal{R}=\mathcal{R}(T)$. On the other hand, the field equations (\ref{Fieldeqs1}) can be expressed just in terms of the spacetime metric, its derivatives and the energy-momentum tensor by using the conformal transformation (\ref{conform_transform22}), under which the Ricci tensor $\mathcal{R}_{\mu\nu}$ yields:
\be
\mathcal{R}_{\mu\nu}(q)=R_{\mu\nu}(g)+\frac{4}{\Omega^2}\nabla_{\mu}\Omega\nabla_{\nu}\Omega-\frac{2}{\Omega}\nabla_{\mu}\nabla_{\nu}\Omega-g_{\mu\nu}\frac{g^{\rho\sigma}}{\Omega^2}\nabla_{\rho}\Omega\nabla_{\sigma}\Omega-g_{\mu\nu}\frac{\Box\Omega}{\Omega}\ .
\label{ConformalR}
\ee
where the covariant derivatives in the rhs of this expression are the Christoffel symbols defined in terms of the spacetime metric $g_{\mu\nu}$. Finally, the field equations (\ref{Fieldeqs1}) are written as:
\be
R_{\mu\nu}(g)-\frac{1}{2}g_{\mu\nu}R(g)=\frac{\kappa^2}{f_{\mathcal{R}}}T_{\mu\nu}-g_{\mu\nu}\frac{\mathcal{R}f_{\mathcal{R}}-f}{2f_{\mathcal{R}}}-\frac{3}{2f_{\mathcal{R}}^2}\left[\nabla_{\mu}f_{\mathcal{R}}\nabla_{\nu}f_{\mathcal{R}}-\frac{1}{2}g_{\nu\mu}\nabla_{\lambda}f_{\mathcal{R}}\nabla^{\lambda}f_{\mathcal{R}}\right]+\frac{1}{f_{\mathcal{R}}}\left[\nabla_{\mu}\nabla_{\nu}f_{\mathcal{R}}-g_{\mu\nu}\Box f_{\mathcal{R}}\right]\ .
\label{fieldEq2}
\ee
Hence, the set of field equations for $f(\mathcal{R})$ gravity within the Palatini formalism, given by (\ref{Fieldeqs1}) and (\ref{Fieldeqs}) are now given by the algebraic equation (\ref{traceRT}) and by the field equations (\ref{fieldEq2}), which are just the Einstein field equations with a non-standard matter side of the equations. Moreover, the field equations (\ref{fieldEq2}) are actually equivalent to the ones of a Brans-Dicke-like theory, as can be easily shown just by identifying $\phi=f_{\mathcal{R}}$ and  $V(\phi)=\mathcal{R}\phi-f(\mathcal{R})$, and by using the conformal transformation of the Ricci tensor (\ref{ConformalR}),  such that the gravitational action  (\ref{fRaction}) is expressed as follows \cite{DeFelice:2010aj}:
\be
S=\frac{1}{2\kappa^2}\int d^4x\sqrt{-g}\left[\phi R(g)+\frac{3}{2\phi}\partial_{\mu}\phi\partial^{\mu}\phi-V(\phi)+L_m\right]\ ,
\label{Brans-DickeEquiv}
\ee
This gravitational action can be easily identified with the one for a Brans-Dicke-like theory with $w=-3/2$ and a potential, which shows that the scalar field is non-dynamical, since the corresponding scalar field equation has its kinetic term missing. Hence, through these tools our aim is to analyse the $3+1$ decomposition in $f(\mathcal{R})$ gravity and to obtain the corresponding Gauss-Codazzi relations, in the next section.
\section{$3+1$ decomposition and Gauss-Codazzi relations}
\label{tresmasuno}

To establish the $3+1$ decomposition of a generic  $f(\mathcal{R})$ gravity theory in the Palatini formalism, the spacetime is conveniently decomposed onto hypersurface(s) and the corresponding Gauss-Codazzi equations are obtained, which allow us to express the curvature tensor, restricted to the hypersurface(s), in terms of the intrinsic and extrinsic curvature of the hypersurfaces. Such hypersurfaces might represent the matching hypersurface between different regions of the spacetime or a domain wall in the braneworld scenario, such that obtaining the corresponding junction conditions is fundamental and is one of the application performed in this paper of the  $3+1$ decomposition within these theories. Hence, here we decompose the spacetime and express the variables that characterises a particular gravitational theory in terms defined on the hypersurface. Let us consider an embedded hypersurface $\Sigma$ defined as a subspace of a $3+1$ dimensional manifold $\mathcal{M}$, as follows:
 \be
  \Sigma=\left\{x\in \mathcal{M}: S(x)=0\right\}\ ,
  \label{hypersurface}
  \ee
where $S(x)$ is a real function that defines the hypersurface. Equivalently, one may think $\Sigma$ as an embedding hypersurface in $\mathcal{M}$, such that the parametric equations among the coordinates defined in $\Sigma$ ($y^a$) and the coordinates of $\mathcal{M}$ ($x^{\mu}$) are:
\be
x^{\mu}=x^{\mu}(y^{a})\ .
\label{parametric}
\ee
We will use greek indexes for referring to the spacetime coordinates/variables and latin ones when referring to coordinates/variables defined on the hypersurface $\Sigma$. The corresponding normal vector to the hypersurface can be expressed as:
 \be
n_{\mu}=\epsilon\frac{\partial_{\mu}S(x)}{\sqrt{|g^{\mu\nu}\partial_{\mu}S\partial_{\nu}S|}}\ .
\label{normalvector}
\ee
Here $\epsilon=\pm 1$, depending on whether the hypersurface is timelike or spacelike. The induced metric $\gamma_{\mu\nu}$ on $\Sigma$ is given by:
\be
\gamma_{\mu\nu}=g_{\mu\nu}-\epsilon n_{\mu}n_{\nu}\ ,
\label{inducedmetric}
\ee
where $g_{\mu\nu}$ is the metric of the spacetime manifold $\mathcal{M}$. This tensor is orthogonal to $n^{\mu}$, and allow us to obtain the tangential components to the hypersurface $\Sigma$ of any tensor. Moreover, by the parametric equations (\ref{parametric}), the corresponding induced metric in $\Sigma$ is given by:
\be
\gamma_{ab}=\frac{\partial x^{\mu}}{\partial y^{a}}\frac{\partial x^{\nu}}{\partial y^{b}}\gamma_{\mu\nu}\ .
\label{inducedmetric2}
\ee
As usual, the extrinsic curvature is defined as the projection of the covariant derivative of the normal vector along the hypersurface,
\be 
K_{\mu\nu}=\gamma^{\alpha}_{\mu}\gamma^{\beta}_{\nu}\nabla_{\alpha}n_{\beta}\ ,
\label{extrinsic}
\ee
which reduces to $K_{\mu\nu}=\nabla_{\mu}n_{\nu}$ for a family of hypersurfaces extended off $\Sigma$ along a geodesic vector field. In addition, the extrinsic curvature (\ref{extrinsic}) can be also expressed in terms of the hypersurface indexes as:
\be
K_{ab}=\frac{\partial x^{\mu}}{\partial y^{a}}\frac{\partial x^{\nu}}{\partial y^{b}}K_{\mu\nu}\ .
\label{extrinsic2}
\ee
Hence, the Ricci scalar, defined by the connection compatible with the spacetime metric, can be expressed in terms of the intrinsic and extrinsic curvature on the hypersurface $\Sigma$ as follows:
\be
R={}^{(3)}R+\epsilon (K_{\mu\nu}K^{\mu\nu}-K^2)+2\epsilon n^{\mu}n^{\nu}R_{\mu\nu}\ ,
\label{RicciScaHyper1}
\ee
where ${}^{(3)}R$ is the three dimensional curvature scalar of the hypersurface. Alternatively, one may substitute the normal projection of the Ricci tensor by using some identities to yield \cite{Lecturenotes}:
\be
R={}^{(3)}R+\epsilon (K^2-K_{\mu\nu}K^{\mu\nu})+2\epsilon \nabla_{\mu}\left(n^{\nu}\nabla_{\nu}n^{\mu}-n^{\mu}\nabla_{\nu}n^{\nu}\right)\ .
\label{RicciScaHyper2}
\ee
This is one of the so-called Gauss-Codazzi relations that eases to construct the ADM formalism in General Relativity and leads naturally to the Gibbons-Hawking-York (GHY) boundary term, as can be easily shown through the Hilbert-Einstein action:
\bea
S_{EH}&=&\int d^4x\sqrt{-g} \left[{}^{(3)}R+\epsilon (K^2-K_{\mu\nu}K^{\mu\nu})+2\epsilon \nabla_{\mu}V^{\mu}\right]= \nn
&=&\int d^4x\sqrt{-g} \left[{}^{(3)}R+\epsilon (K^2-K_{\mu\nu}K^{\mu\nu})\right]+2\int d^3y\sqrt{-\gamma} n_{\mu}V^{\mu} ,
\label{EHaction}
\eea
where $K=\nabla_\nu n^{\nu}$ and $V^{\mu}$ is given by:
\be
V^{\mu}=n^{\nu}\nabla_{\nu}n^{\mu}-n^{\mu}\nabla_{\nu}n^{\nu}\ .
\ee
And we have used the Gauss-Sokes theorem on the total derivative:
\be
\int_{\mathcal{M}} d^4x \sqrt{-g}\nabla_{\sigma}V^{\sigma}=\epsilon\int_{\Sigma} d^3y \sqrt{|\gamma|} n_{\sigma}V^{\sigma}\ .
\label{BSTheorem}
\ee
After some manipulations, the boundary term in (\ref{EHaction}) can be expressed as:
\be
2 \int d^3y\sqrt{-\gamma} n_{\mu}\left(n^{\nu}\nabla_{\nu}n^{\mu}-n^{\mu}\nabla_{\nu}n^{\nu}\right)=-2\epsilon \int d^3y\sqrt{-\gamma} K\ .
\label{GYHterm}
\ee
This is the so-called GHY term that should be subtracted to the Hilbert-Einstein action to have a well-posed variational principle with standard Dirichlet boundary conditions on the variations of the metric, such that the so-called Gauss-Codazzi form of the gravitational action for GR is obtained:
\be
S=S_{EH}+S_{GC}=\int d^4x\sqrt{-g} R+2\epsilon \int d^3x\sqrt{-\gamma}\ K=\int d^4x\sqrt{-g} \left[{}^{(3)}R+\epsilon (K^2-K_{\mu\nu}K^{\mu\nu})\right]\ .
\label{GCGR}
\ee
Let us now extend this formalism to $f(R)$ gravities. To do so, we start by obtaining the corresponding Gauss-Codazzi action for metric $f(R)$ gravity, where the connection is assumed to be compatible with the metric. A simple way to proceed lie in expressing the $f(R)$ action in terms of its equivalence scalar-tensor form:
\be
S=\int d^4x\sqrt{-g} f(R)=\int d^4x\sqrt{-g} \left[\phi R-U(\phi)\right]\ .
\label{FR_scalarTensor}
\ee
By varying the action with respect to the scalar field, the mapping among both actions is easily obtained:
\be
\phi=f_R\ , \quad U(\phi)=Rf_R-f\ .
\label{mapping}
\ee
Hence, just by using the relation (\ref{RicciScaHyper2}), the action (\ref{FR_scalarTensor}) turns out:
\be
S=\int d^4x\sqrt{-g} \left[\phi \left({}^{(3)}R+\epsilon (K^2-K_{\mu\nu}K^{\mu\nu})+2\epsilon \nabla_{\mu}V^{\mu}\right)-U(\phi)\right]\ .
\label{FR_Boundary_term}
\ee
Integrating by parts the term $ \nabla_{\mu}V^{\mu}$, the action (\ref{FR_Boundary_term}) yields:
\bea
S&=&\int d^4x\sqrt{-g} \left[\phi \left({}^{(3)}R+\epsilon (K^2-K_{\mu\nu}K^{\mu\nu})\right)-2\epsilon V^{\mu} \nabla_{\mu}\phi-U(\phi)+2\epsilon \nabla_{\mu}\left(V^{\mu}\phi\right)\right]=\nn
&=&\int d^4x\sqrt{-g} \left[\phi \left({}^{(3)}R+\epsilon (K^2-K_{\mu\nu}K^{\mu\nu})\right)-2\epsilon V^{\mu} \nabla_{\mu}\phi-U(\phi)\right]+2\int d^3y\sqrt{\gamma}n_{\mu}V^{\mu}\phi\ .
\label{FR_Boundary_term2}
\eea
And the boundary term can be expressed as follows:
\be
2\int d^3y\sqrt{\gamma}\ \phi\ n_{\mu}V^{\mu}=-2\epsilon\int d^3y\sqrt{\gamma}\phi\ K=-2\epsilon\int d^3y\sqrt{\gamma}\ f_R\ K\ .
\label{GHY_FR}
\ee
This is the analog to the GBY in metric $f(R)$ gravities, as shown in \cite{Madsen:1989rz,Nojiri:2000kh,Dyer:2008hb,Guarnizo:2010xr} by variational principles. Then, the Gauss-Codazzi action for metric $f(R)$ gravity leads to:
\be
S=\int d^4x\sqrt{-g} f(R)+2\epsilon\int d^3y\sqrt{\gamma}\ f_R\ K= \int d^4x\sqrt{-g} \left[\phi\left({}^{(3)}R+\epsilon (K^2-K_{\mu\nu}K^{\mu\nu})\right)-2\epsilon V^{\mu} \nabla_{\mu}\phi-U(\phi)\right]\ .
\label{FR_GC_action}
\ee
Note that by varying the action with respect to the scalar field, the mapping (\ref{mapping}) is recovered. \\

The case of modified gravities in the Palatini formalism is a bit more tricky, as one intends to express the Ricci scalar (\ref{Ricci}) in terms of the intrinsic and extrinsic curvature of the spacetime hypersurface and the independent connection does not depend in principle on the spacetime metric. Nevertheless, the equation (\ref{hmetricity}) states that the connection is compatible with the conformal metric $q_{\mu\nu}$, which is related to the spacetime metric $g_{\mu\nu}$ by the conformal transformation (\ref{conform_transform22}. Then, the normal vector and the induced metric on the hypersurface transform as (see Appendix C in Ref.~\cite{Dyer:2008hb}):
\be
\tilde{n}_{\mu} =\Omega\ n_{\mu}\ , \quad \tilde{\gamma}_{\mu\nu}=\Omega^2  \gamma_{\mu\nu}=q_{\mu\nu}-\epsilon \tilde{n}_{\mu} \tilde{n}_{\nu} \ ,
\label{conform_normal_induced}
\ee
where recall that $q_{\mu\nu}=\Omega^2 g_{\mu\nu}$ with $\Omega^2=f_{\mathcal{R}}$, whereas the extrinsic curvature defined in terms of the connection compatible to $q_{\mu\nu}$ yields:
\be
\mathcal{K}_{\mu\nu}=\Omega K_{\mu\nu}+\gamma_{\mu\nu}n^{\alpha}\partial_{\alpha}\Omega=\tilde{\nabla}_{\mu}\tilde{n}_{\nu}\ .
\label{extrinsicConform}
\ee
Alternatively to the Brans-Dicke equivalent action (\ref{Brans-DickeEquiv}), we can express the action (\ref{fRaction}) for the Palatini formalism similarly to the $f(R)$ metric case as follows:
\be
S=\int d^4x\sqrt{-g} f(\mathcal{R})=\int d^4x\sqrt{-g} \left[\phi \mathcal{R}-U(\phi)\right]\ .
\label{FR_scalarTensor2}
\ee
The relation among both actions is analog to the metric case (\ref{mapping}), i.e. $\phi=f_{\mathcal{R}}$ and $U=\mathcal{R}f_{\mathcal{R}}-f(\mathcal{R})$, being $\phi$ an auxiliary scalar field with no dynamics. By applying the conformal transformation (\ref{conform_transform22}) to the spacetime metric, the Ricci scalar $\mathcal{R}$ yields:  
\be
\mathcal{R}=g^{\mu\nu}\mathcal{R}_{\mu\nu}=\Omega^2 q^{\mu\nu}\mathcal{R}_{\mu\nu}\ ,
\ee
while the action (\ref{FR_scalarTensor2}) is transformed as:
\be
S=\int d^4x\sqrt{-q} \left[\tilde{\mathcal{R}}(q)-\frac{U(\phi)}{\phi^2}\right]\ ,
\label{FR_scalarTensor3}
\ee
where $\tilde{\mathcal{R}}(q)=q^{\mu\nu}\mathcal{R}_{\mu\nu}$ is the contraction of the Ricci tensor with the conformal metric $q_{\mu\nu}$. Note that now the gravitational action depends solely on the metric $q_{\mu\nu}$ and not on the spacetime metric $g_{\mu\nu}$. Hence, the action for $f(\mathcal{R})$ in the Palatini formalism reduces to the Hilbert-Einstein action with a cosmological constant in vacuum, equivalently to the field equations (\ref{Fieldeqs1}) and (\ref{Fieldeqs}). 
Hence, by the induced metric $\tilde{\gamma}_{\mu\nu}$ and the normal vector $\tilde{n}_{\mu}$ given in (\ref{conform_normal_induced}), the Ricci scalar $\tilde{\mathcal{R}}$ can be expressed in terms of the extrinsic curvature $\mathcal{K}$ and the intrinsic curvature ${}^{(3)}\tilde{\mathcal{R}}$ conformally related to the ones of the spacetime hypersurface, as follows:
\bea
\tilde{\mathcal{R}}&=& q^{\mu\nu}\mathcal{R}_{\mu\nu}=\Omega^{-2}\ \mathcal{R}\nn
&=&\left[{}^{(3)}\tilde{\mathcal{R}}+\epsilon (\mathcal{K}^2-\mathcal{K}_{\mu\nu}\mathcal{K}^{\mu\nu})+2\epsilon \tilde{\nabla}_{\mu}\left(\tilde{n}^{\nu}\tilde{\nabla}_{\nu}\tilde{n}^{\mu}-\tilde{n}^{\mu}\tilde{\nabla}_{\nu}\tilde{n}^{\nu}\right)\right]\ .
\label{Ricci_Pala_descom}
\eea
This establishes the corresponding Gauss-Codazzi relation for the curvature scalar (\ref{Ricci}) in the Palatini formalism. Let us now obtain the boundary term in $f(\mathcal{R})$ gravity and the Gauss-Codazzi action by using the relation (\ref{Ricci_Pala_descom}). By substituting the relation (\ref{Ricci_Pala_descom}) into the action (\ref{FR_scalarTensor3}), we get:
\be
S=\int d^4x\sqrt{-q} \left[{}^{(3)}\tilde{\mathcal{R}}+\epsilon (\mathcal{K}^2-\mathcal{K}_{\mu\nu}\mathcal{K}^{\mu\nu})+2\epsilon \tilde{\nabla}_{\mu}\mathcal{V}^{\mu}-\frac{U(\phi)}{\phi^2}\right]\ .
\label{FR_Boundary_termP}
\ee
where $\mathcal{V}^{\nu}=\tilde{n}^{\nu}\tilde{\nabla}_{\nu}\tilde{n}^{\mu}-\tilde{n}^{\mu}\tilde{\nabla}_{\nu}\tilde{n}^{\nu}$. As in the case of the Hilbert-Einstein action, we can apply directly the Gauss-Stokes theorem on the total derivative, leading to:
\be
2\epsilon\int d^4x\tilde{\nabla}_{\mu}\left(\sqrt{-q}\mathcal{V}^{\mu}\right)=2\int d^3y\sqrt{\tilde{\gamma}}\tilde{n}_{\mu}\mathcal{V}^{\mu}=-2\epsilon\int d^3y\sqrt{\tilde{\gamma}}\mathcal{K}\ .
\label{Boundary_term_palatini}
\ee
This is the boundary term that has to be subtracted to the Palatini action which coincide with the one found in Ref.~\cite{Obukhov1987} for the Hilbert-Einstein action and in Ref.~\cite{Gomez:2020rnq} for $f(\mathcal{R})$ gravity by following variational principles. Then, the Gauss-Codazzi action for $f(\mathcal{R})$ gravity in the Palatini formalism yields:
\be
S=\int d^4x\sqrt{-g} f(\mathcal{R})+2\epsilon\int d^3y\sqrt{\tilde{\gamma}}\mathcal{K}= \int d^4x\sqrt{-q} \left[{}^{(3)}\tilde{\mathcal{R}}+\epsilon (\mathcal{K}^2-\mathcal{K}_{\mu\nu}\mathcal{K}^{\mu\nu})-\frac{U(\phi)}{\phi^2}\right]\ .
\label{FR_Palatini_GC_action}
\ee
Hence, the Gauss-Codazzi action in the Palatini formalism is equivalent to the one in GR with a cosmological constant. Nevertheless, in the presence of a matter Lagrangian, the conformal transformation (\ref{conform_transform22}) implies a coupling among the scalar field and matter, as the action (\ref{fRaction}) is transformed as:
\be
S=\int d^4x\sqrt{-q} \left[\tilde{\mathcal{R}}(q)-\frac{U(\phi)}{\phi^2}+\frac{1}{\phi^2}L_m\right]\ .
\label{FR_scalarTensor4}
\ee
The variation of the action with respect to the conformal metric $q_{\mu\nu}$ results in the following field equations:
\be
\tilde{\mathcal{R}}_{\mu\nu}-\frac{1}{2}q_{\mu\nu}\tilde{\mathcal{R}}+\frac{1}{2}q_{\mu\nu}\frac{U(\phi)}{\phi^2}=\frac{\kappa^2}{\phi^2}\tilde{T}_{\mu\nu},
\label{field_equations_conf}
\ee
where $\tilde{T}_{\mu\nu}=-\frac{2}{\sqrt{-q}}\frac{\delta \tilde{S}_m}{\delta q^{\mu\nu}}=\Omega^2T_{\mu\nu}$ is the energy-momentum tensor defined in terms of the variation of the matter action with respect to $q_{\mu\nu}$. While the variation of the action with respect to the scalar field $\phi$ provides the complementary constraint equation:
\be
2U(\phi)-\phi U'(\phi)=\kappa^2 T\ ,
\label{Scalar_field_eq11}
\ee
which is equivalent to the trace equation (\ref{traceRT}). Hence, we have reformulated the gravitational action for $f(\mathcal{R})$ in the Palatini formalism equivalently to a Brans-Dicke theory expressed in the Einstein frame through the conformal transformation (\ref{conform_transform22}), such that the corresponding $3+1$ decomposition is easily achieved and the Gauss-Codazzi relations are obtained. In the next sections, we apply such decomposition to two well known and fundamental frameworks, the ADM decomposition and the junction conditions.

\section{ADM decomposition}
\label{ADM}

The most direct application of the Gauss-Codazzi relations obtained above is the so-called ADM decomposition. The ADM decomposition consists in a foliation of the spacetime onto hypersurfaces of constant time, such that the family of normal vectors to the hypersurfaces is given by:
\be
n_{\mu}=-N \partial_{\mu} t\ ,
\ee
where $N$ is the so-called lapse function:
\be
N=\frac{1}{\sqrt{-g^{\mu\nu}\partial_{\mu}t\ \partial_{\nu}t}}\ .
\ee
The spacetime metric can be expressed in terms of the ADM variables as \cite{Gourgoulhon:2007ue}:
\be
ds^2=g_{\mu\nu}dx^{\mu}dx^{nu}=-N^2dt^2+\gamma_{ij}\left(dx^{i}+N^{i}dt\right)\left(dx^{j}+N^{j}dt\right)\ ,
\label{metricADM}
\ee
where $N^{i}$ is the shift vector and $\gamma_{ij}$ is the first fundamental form or induced metric on the hypersurface of constant time:
\be
\gamma_{\alpha\beta}=g_{\alpha\beta}+n_{\alpha}n_{\beta}\ .
\label{ADM_ind_metric}
\ee
Whereas the extrinsic curvature is given by \footnote{To keep the usual convention in the ADM formalism, in section \ref{ADM} we have used $K_{ij}=-\gamma^{\alpha}_{\mu}\gamma^{\beta}_{\nu}\nabla_{\alpha}n_{\beta}$ with a minus sign instead of (\ref{extrinsic}).}:
\be
K_{ij}=-\frac{1}{N}\left(\partial_{t}\gamma_{ij}-D_{i}N_{j}-D_{j}N_{i}\right)\ ,
\ee
where the covariant derivatives $D_i$ are associated to the induced metric $\gamma_{ij}$ on the hypersurface $\Sigma_t$. By using the Gauss-Codazzi equations, the scalar curvature can be written in terms of the intrinsic and extrinsic curvatures as follows:
\be
R={}^{(3)}R+K^2-K_{ij}K^{ij}-\frac{2}{N}\left(\partial_t K-N^{i}\partial_{i}K\right)-\frac{2}{N}D_{i}D^{i}N\ ,
\label{Ricci_scalar_ADM}
\ee
To extend this decomposition to Palatini $f(\mathcal{R})$ gravity, we can proceed similarly as in the previous section and use the Gauss-Codazzi equations obtained above to express the scalar curvature in terms of the ADM variables. To do so, we use the conformal transformation (\ref{conform_transform22}) that relates the metric $q_{\mu\nu}$, compatible with the connection that defines the Ricci tensor $\mathcal{R}_{\mu\nu}$, with the spacetime metric $g_{\mu\nu}$, such that the line element transformed as:
\be
d\tilde{s}^2=\Omega^2 ds^2=-\mathcal{N}^2dt^2+\tilde{\gamma}_{ij}\left(dx^{i}+N^{i}dt\right)\left(dx^{j}+N^{j}dt\right)\ ,
\label{metricADMconfor}
\ee
where we have defined:
\be
\mathcal{N}=\Omega N\ , \quad \tilde{\gamma}_{ij}=\Omega^2\gamma_{ij}\ .
\ee
While the conformal metric yields:
\be
\tilde{\gamma}_{\alpha\beta}=q_{\alpha\beta}+\tilde{n}_{\alpha}\tilde{n}_{\beta}\ .
\label{ADM_ind_metric_Conf}
\ee
And the extrinsic curvature (\ref{extrinsicConform}) is given by:
\be
\mathcal{K}_{ij}=-\frac{1}{\mathcal{N}}\left(\partial_{t}\tilde{\gamma}_{ij}-\mathcal{D}_{i}N_{j}-\mathcal{D}_{j}N_{i}\right)\ ,
\ee
Here the spatial covariant derivatives $\mathcal{D}_i$ are compatible to the conformal metric $\tilde{\gamma}_{ij}$. Finally, by using the corresponding Gauss-Codazzi relation given in (\ref{Ricci_Pala_descom}), the Ricci scalar defined as the contraction of the Ricci tensor $\mathcal{R}_{\mu\nu}$ with $q_{\mu\nu}$ is obtained:
\be
\tilde{\mathcal{R}}={}^{(3)}\tilde{\mathcal{R}}+\mathcal{K}^2-\mathcal{K}_{ij}\mathcal{K}^{ij}-\frac{2}{\mathcal{N}}\left(\partial_t \mathcal{K}-N^{i}\partial_{i}\mathcal{K}\right)-\frac{2}{\mathcal{N}}\mathcal{D}_{i}\mathcal{D}^{i}\mathcal{N}\ ,
\label{Ricci_scalar_ADM_Palatini}
\ee
which states the way to apply the ADM decomposition to the action and field equations in $f(\mathcal{R})$ gravity. Alternatively, one can proceed with this decomposition directly from the equivalent action and field equations expressed as a Brans-Dicke-like theory, as given in (\ref{fieldEq2}) and (\ref{Brans-DickeEquiv}), where the gravitational action and the equations are expressed just in terms of the spacetime metric and a scalar field that is coupled to the energy-momentum tensor. In the next section, we will use such picture to obtain explicitly the junction conditions for this class of gravitational theories.

\section{Junction conditions}
\label{Junctions}


In a previous and recent paper \cite{Olmo:2020fri}, the corresponding junction conditions for this class of theories were obtained by expressing the field equations in terms of distributions, an extension of the approach studied previously in metric $f(R)$ gravity \cite{Senovilla:2013vra} and recently in \cite{Chu:2021uec}. Here we apply the $3+1$ decomposition, as done before in metric $f(R)$ theories \cite{Deruelle:2007pt} or in some extensions of Teleparallel gravity \cite{delaCruz-Dombriz:2014zaa}, to obtain the corresponding junction conditions in Palatini $f(\mathcal{R})$ gravity which coincide with the ones calculated in \cite{Olmo:2020fri}. To do so, the metric tensor is expressed in the so-called Gaussian-normal coordinates and then through the Gauss-Codazzi relations the matching conditions between two different regions of the spacetime are obtained. The spacetime metric in  Gaussian-normal coordinates is given by:
\be
{\rm d}s^2={\rm d}y^2+\gamma_{ij}{\rm d}x^i{\rm d}x^j\ .
\label{Gaussian_normal}
\ee
Hence, the boundary/brane between different spacetime regions is located at $y=0$, whereas $\gamma_{ij}$ is the induced metric on the matching hypersurface. The corresponding normal vector is given by:
\be
n_{\mu}=\partial_{\mu}y\ .
\label{normal_vactor_GC}
\ee
While the extrinsic curvature (\ref{extrinsic}) of the hypersurface is easily obtained,
\be
K_{ij}=\frac{1}{2}\partial_y\gamma_{ij}\ .
\label{extrinsicGN}
\ee
We can now proceed within two different but equivalent approaches: by applying the conformal transformation (\ref{conform_transform22}) and expressing all the quantities in terms of the $q_{\mu\nu}$ tensor, i.e. the field equations (\ref{field_equations_conf}, \ref{Scalar_field_eq11}), or alternatively through the field equations expressed in terms of the spacetime metric $g_{\mu\nu}$ and the trace of the energy-momentum tensor (\ref{fieldEq2}). Both approaches lead to the same conditions but the latter simplifies the calculations, as the hypersurface is defined through the spacetime metric instead of the conformal metric. Hence, let us deduce the junction conditions through the field equations (\ref{fieldEq2}) and then, will be expressed in terms of the variables defined by the conformal metric. To do so, the components of the Ricci tensor $R_{\mu\nu}(g)$ can be expressed in terms of the induced metric and the extrinsic curvature as follows \cite{Deruelle:2007pt}:
\bea
R_{yy}=-\gamma^{ij}\frac{\partial K_{ij}}{\partial y}+K_{ij}K^{ij}\ , \quad R_{yj}=D_iK^{i}_{j}-D_{j}K\ ,\nn
R_{ij}={}^{(3)}R_{ij}-\frac{\partial K_{ij}}{\partial y}+K_{ki}K^{k}_{j}+\gamma^{kl}K_{ik}K_{lj}-K K_{ij}\ ,
\eea
while the Ricci scalar is given by:
\be
R=-2\frac{\partial K}{\partial y}-K_{ij}K^{ij}-K^2+{}^{(3)}R\ .
\ee
Hence, the corresponding decomposition of the Einstein tensor yields:
\bea
G_{yy}=-\frac{1}{2}\left(K_{ij}K^{ij}-K^2+{}^{(3)}R\right)\ , \quad G_{yi}=D_j\left(K^{j}_i-\delta^{j}_{i}K\right)\ , \nn
G_{ij}=\partial_{y}\left(\gamma_{ij}K-K_{ij}\right)+2K^{k}_{i}K_{kj}-3KK_{ij}+\frac{1}{2}\gamma_{ij}\left(K_{kl}K^{kl}+K^2\right)+{}^{(3)}G_{ij}\ .
\label{Einstein_Ten1}
\eea
As in the GR case, the first junction condition follows directly from (\ref{extrinsicGN}) in order to avoid powers of Dirac's delta distributions as $\delta^2(y)$ in the Einstein tensor (\ref{Einstein_Ten1}) and consequently in the field equations, such that the induced metric $\gamma_{ij}$ has to be continuous across the boundary/brane located at $y=0$:
\be
\left[ \gamma_{ij} \right]^{+}_{-}=0\ .
\label{Firstjunction}
\ee
The extrinsic curvature or second fundamental form (\ref{extrinsicGN}) might be discontinuous, as contains first derivatives of the induced metric, and enters in the field equations through the $ij$-component of the Einstein tensor (\ref{Einstein_Ten1}), leading to:
\be
\partial_{y}\left(\gamma_{ij}K-K_{ij}\right)=P_{ij}\delta(y)\ ,
\label{ExtrinsDelta}
\ee
Integrating across the boundary, we obtain:
\be
P_{ij}=\left[\gamma_{ij}K-K_{ij}\right]^{+}_{-}\ .
\ee
In absence of $\delta$'s in the rhs of the field equations, one leads to:
\be
\left[K_{ij}\right]^{+}_{-}=0\ ,
\ee
which imposes continuity on the extrinsic curvature or second fundamental form $K_{ij}$. This is obviously the case of GR, nothing surprising as the lhs of the field equations as written in (\ref{fieldEq2}) is the same as in GR. Let us now analyse the rhs of (\ref{fieldEq2}), which is given by:
\be
\frac{\kappa^2}{f_{\mathcal{R}}}T_{\mu\nu}-g_{\mu\nu}\frac{\mathcal{R}f_{\mathcal{R}}-f}{2f_{\mathcal{R}}}-\frac{3}{2f_{\mathcal{R}}^2}\left[\nabla_{\mu}f_{\mathcal{R}}\nabla_{\nu}f_{\mathcal{R}}-\frac{1}{2}g_{\nu\mu}\nabla_{\lambda}f_{\mathcal{R}}\nabla^{\lambda}f_{\mathcal{R}}\right]+\frac{1}{f_{\mathcal{R}}}\left[\nabla_{\mu}\nabla_{\nu}f_{\mathcal{R}}-g_{\mu\nu}\Box f_{\mathcal{R}}\right]\ .
\label{rhsFieldEQ}
\ee
From the trace equation (\ref{traceRT}), we have $\mathcal{R}=\mathcal{R}(T)$, such that any derivative in (\ref{rhsFieldEQ}) reduces to derivatives on the trace of the energy-momentum tensor. Given the terms as $\nabla_{\mu}f_{\mathcal{R}}\nabla_{\nu}f_{\mathcal{R}}$ that lead to products of the type $\partial_y T\partial_y T$, such that one requires to remove any delta function from the trace of the energy-momentum tensor and its first derivative what leads to impose continuity on its trace, reaching to the second junction condition \cite{Olmo:2020fri}:
\be
\left[T\right]^{+}_{-}=0\ .
\label{SecondJunc}
\ee
Note that this is based on the assumption that equation (\ref{traceRT}) has real solutions given by $\mathcal{R}=\mathcal{R}(T)$, which additionally might be defined just in a particular domain. Otherwise,  In addition, from the $ij-$component of (\ref{rhsFieldEQ}), the D'Alambertien $\Box f_{\mathcal{R}}$ contains second derivatives of the trace of the energy-momentum tensor that might introduce delta distributions through the term:
\be
-\frac{1}{f_{\mathcal{R}}}\gamma_{ij}\Box f_{\mathcal{R}}\ \quad \rightarrow \quad \partial_{y}\left[-\frac{1}{f_{\mathcal{R}}}\gamma_{ij}\partial_{y}f_{\mathcal{R}}\right]=S_{ij}\delta(y)\ .
\ee
Integrating across the boundary, it yields:
\be
S_{ij}=\left[-\frac{1}{f_{\mathcal{R}}}\gamma_{ij}\partial_{y}f_{\mathcal{R}}\right]^{+}_{-}=-\frac{1}{f_{\mathcal{R}}}\gamma_{ij}\left[\partial_{y}f_{\mathcal{R}}\right]^{+}_{-}\ .
\ee
In order to cancel the deltas in both sides of the field equations, we have to compensate this term with (\ref{ExtrinsDelta}) and also with the possible divergences of the energy-momentum tensor $\tau_{ij}$, since the condition (\ref{SecondJunc}) just imposes continuity on the trace of the energy-momentum tensor, such that the third junction condition leads to:
\be
\left[\gamma_{ij}K-K_{ij}\right]^{+}_{-}=\frac{\kappa^2}{f_{\mathcal{R}}}\tau_{ij}-\frac{1}{f_{\mathcal{R}}}\gamma_{ij}\left[\partial_{y}f_{\mathcal{R}}\right]^{+}_{-}\ .
\label{ThirdJunc1}
\ee
However, note that due to the presence of terms as $\partial_y T\partial_y T$ in the field equations, besides the second junction condition (\ref{SecondJunc}), one has to impose that the trace of the singular part the energy-momentum tensor becomes null:
\be
\tau=\tau^{\mu}_{\;\;\mu}=0\ .
\label{traceSing}
\ee
And the trace of the third junction condition (\ref{ThirdJunc1}) yields:
\be
\left[K\right]^{+}_{-}=-\frac{3}{2}\frac{1}{f_{\mathcal{R}}}\left[\partial_{y}f_{\mathcal{R}}\right]^{+}_{-}=-\frac{3}{2}\frac{f_{\mathcal{R}\mathcal{R}}}{f_{\mathcal{R}}}R_T \left[\partial_{y} T\right]^{+}_{-}\ .
\label{ThirdJunc2}
\ee
Hence, the set given by (\ref{Firstjunction}), (\ref{SecondJunc}) and (\ref{ThirdJunc2}) form the junction conditions for $f(\mathcal{R})$ gravity in the Palatini formalism, which coincide with the ones found in \cite{Olmo:2020fri} through a distributional analysis. Note that these conditions might be found through the conformal frame defined by the conformal transformation (\ref{conform_transform22}). Firstly, since the transformation of the extrinsic curvature (\ref{extrinsicConform}) contains derivatives of $\Omega^2=f_{\mathcal{R}}(\mathcal{R}(T))$ and consequently of the trace of the energy-momentum tensor, such that the second junction condition (\ref{SecondJunc}) is automatically achieved. Then, this also implies the continuity of the induced conformal metric:
\be
\left[ \tilde{\gamma}_{ij} \right]^{+}_{-}=0\ .
\label{Firstjunctiona}
\ee
Whereas the third junction condition (\ref{ThirdJunc2}) can be rewritten in terms of the conformal extrinsic curvature by using the transformations (\ref{conform_transform22}), such that in absence of divergences in the energy-momentum tensor, the third junction condition (\ref{ThirdJunc1}) yields:
\be
\left[\tilde{\gamma_{ij}}\mathcal{K}-\mathcal{K}_{ij}\right]^{+}_{-}=0\ .
\label{ThirdJunc1a}
\ee
which states basically the continuity of the conformal extrinsic curvature:
 \be
\left[\mathcal{K}_{ij}\right]^{+}_{-}=0\ .
\label{ThirdJunc1b}
\ee
Nevertheless, as far as the energy-momentum tensor contains a singular part $\tau_{\mu\nu}$, the third condition (\ref{ThirdJunc1a}) turns out:
\be
\left[\tilde{\gamma_{ij}}\mathcal{K}-\mathcal{K}_{ij}\right]^{+}_{-}=\frac{\kappa^2}{\Omega^3}\tilde{\tau}_{ij}\ ,
\label{ThirdJunc1c}
\ee
where $\tilde{\tau}_{ij}=\Omega^2\tau_{ij}$ and its trace satisfies (\ref{traceSing}). Hence, we have obtained the junction conditions for $f(\mathcal{R})$ gravity in the Palatini formalism, which reduces actually to the Israel ones when are expressed in the conformal variables that define the compatibility of the connection.

\section{Summary}
\label{Conclusions}

Along this manuscript, we have constructed the framework for expressing the corresponding variables that describe $f(\mathcal{R})$ gravity within the Palatini approach in terms of the extrinsic and intrinsic curvatures defined on a hypersurface through the so-called $3+1$ decomposition. To do it so, one takes the advantage of the field equation that establishes the compatibility of the -in principle- independent connection with a metric that is conformally related to the spacetime one, which enables to obtain the conformal related normal vector, extrinsic and intrinsic curvatures for leading to the Gauss-Codazzi relations for the curvature $\mathcal{R}$. By doing so, we have obtained firstly the corresponding Gauss-Codazzi relation in metric $f(R)$ theories and then in Palatini $f(\mathcal{R})$ gravities, where the boundary term that has to be subtracted to the gravitational action is obtained, leading to the Gauss-Codazzi action for both classes of theories, metric $f(R)$ gravities and Palatini $f(\mathcal{R})$ models. This has also a direct application to the so-called ADM decomposition of the spacetime, a foliation of this one onto hypersurfaces of constant time, which has been obtained for these modified gravity theories, enabling the necessary tools for performing a full analysis of the Hamiltonian formulation of the theory.\\

Finally, the $3+1$ decomposition is applied for establishing the corresponding junction conditions in $f(\mathcal{R})$ within the Palatini formalism, which agree with previous results where other approaches were followed \cite{Olmo:2020fri}. The set of matching conditions coincide in vacuum with the ones given in GR but differ in the presence of matter. Nevertheless, when expressing such conditions in terms of the conformal variables, these ones reduce again to the Israel junction conditions in GR. In comparison to metric $f(R)$ theories, where the Ricci scalar and its first derivative has to be continuos in general \cite{Senovilla:2013vra}, and in some particular cases double layers might be allowed \cite{Senovilla:2014kua,Reina:2015gxa}, this not the case for Palatini $f(\mathcal{R})$ gravity, as naturally arise since the field equations remain second order in comparison to the fourth order of metric $f(R)$ gravities. Nevertheless, the theory imposes the continuity of the trace of the energy-momentum tensor which might have important consequences, as was pointed out in \cite{Guerrero:2021pxt}.\\

Hence, the $3+1$ decomposition and Gauss-Codazzi relations obtained here will have important applications and implications on the development and research on modified gravity theories within the Palatini formalism.

\section*{Acknowledgments}
I would like to thank Dr. Diego Rubiera-Garc\'ia for valuable comments on this manuscript. DS-CG is funded by the University of Valladolid (Spain) Ref. POSTDOC UVA20 and by Ministerio de Ciencia e Innovaci\'on (Spain), project Ref. PID2020-117301GA-I00.


\begin{thebibliography}{}


\bibitem{Aldrovandi:2013wha}
R.~Aldrovandi and J.~G.~Pereira,
doi:10.1007/978-94-007-5143-9

\bibitem{BeltranJimenez:2017tkd}
J.~Beltr\'an Jim\'enez, L.~Heisenberg and T.~Koivisto,
Phys. Rev. D \textbf{98}, no.4, 044048 (2018)
doi:10.1103/PhysRevD.98.044048
[arXiv:1710.03116 [gr-qc]].

\bibitem{Olmo:2011uz}
G.~J.~Olmo,
Int. J. Mod. Phys. D \textbf{20}, 413-462 (2011)
doi:10.1142/S0218271811018925
[arXiv:1101.3864 [gr-qc]].

\bibitem{DeFelice:2010aj}
A.~De Felice and S.~Tsujikawa,
Living Rev. Rel. \textbf{13}, 3 (2010)
doi:10.12942/lrr-2010-3
[arXiv:1002.4928 [gr-qc]].

\bibitem{Nojiri:2017ncd}
S.~Nojiri, S.~D.~Odintsov and V.~K.~Oikonomou,
Phys. Rept. \textbf{692}, 1-104 (2017)
doi:10.1016/j.physrep.2017.06.001
[arXiv:1705.11098 [gr-qc]].

\bibitem{BeltranJimenez:2017doy}
J.~Beltran Jimenez, L.~Heisenberg, G.~J.~Olmo and D.~Rubiera-Garcia,
Phys. Rept. \textbf{727}, 1-129 (2018)
doi:10.1016/j.physrep.2017.11.001
[arXiv:1704.03351 [gr-qc]].

\bibitem{Vitagliano:2010sr}
V.~Vitagliano, T.~P.~Sotiriou and S.~Liberati,
Annals Phys. \textbf{326}, 1259-1273 (2011)
[erratum: Annals Phys. \textbf{329}, 186-187 (2013)]
doi:10.1016/j.aop.2011.02.008
[arXiv:1008.0171 [gr-qc]].


\bibitem{Baghram:2009we}
S.~Baghram and S.~Rahvar,
Phys. Rev. D \textbf{80}, 124049 (2009)
doi:10.1103/PhysRevD.80.124049
[arXiv:0912.2410 [astro-ph.CO]];
\bibitem{Aoki:2018lwx}
K.~Aoki and K.~Shimada,
Phys. Rev. D \textbf{98}, no.4, 044038 (2018)
doi:10.1103/PhysRevD.98.044038
[arXiv:1806.02589 [gr-qc]];
\bibitem{Leanizbarrutia:2017xyd}
I.~Leanizbarrutia, F.~S.~N.~Lobo and D.~Saez-Gomez,
Phys. Rev. D \textbf{95}, no.8, 084046 (2017)
doi:10.1103/PhysRevD.95.084046
[arXiv:1701.08980 [gr-qc]];
\bibitem{Rosa:2017jld}
J.~L.~Rosa, S.~Carloni, J.~P.~d.~Lemos and F.~S.~N.~Lobo,
Phys. Rev. D \textbf{95}, no.12, 124035 (2017)
doi:10.1103/PhysRevD.95.124035
[arXiv:1703.03335 [gr-qc]].
\bibitem{Rosa:2019ejh}
J.~L.~Rosa, S.~Carloni and J.~P.~S.~Lemos,
Phys. Rev. D \textbf{101}, no.10, 104056 (2020)
doi:10.1103/PhysRevD.101.104056
[arXiv:1908.07778 [gr-qc]].
\bibitem{Rosa:2021ish}
J.~L.~Rosa, F.~S.~N.~Lobo and D.~Rubiera-Garcia,
[arXiv:2103.02580 [gr-qc]].
\bibitem{Harko:2011nh}
T.~Harko, T.~S.~Koivisto, F.~S.~N.~Lobo and G.~J.~Olmo,
Phys. Rev. D \textbf{85}, 084016 (2012)
doi:10.1103/PhysRevD.85.084016
[arXiv:1110.1049 [gr-qc]];


\bibitem{Shimada:2018lnm}
F.~Bauer and D.~A.~Demir,
Phys. Lett. B \textbf{665}, 222-226 (2008)
doi:10.1016/j.physletb.2008.06.014
[arXiv:0803.2664 [hep-ph]];
\bibitem{Gialamas:2020snr}
I.~D.~Gialamas, A.~Karam and A.~Racioppi,
[arXiv:2006.09124 [gr-qc]].
\bibitem{Antoniadis:2018ywb}
I.~Antoniadis, A.~Karam, A.~Lykkas and K.~Tamvakis,
JCAP \textbf{11}, 028 (2018)
doi:10.1088/1475-7516/2018/11/028
[arXiv:1810.10418 [gr-qc]];
\bibitem{Enckell:2018hmo}
V.~M.~Enckell, K.~Enqvist, S.~Rasanen and L.~P.~Wahlman,
JCAP \textbf{02}, 022 (2019)
doi:10.1088/1475-7516/2019/02/022
[arXiv:1810.05536 [gr-qc]];
\bibitem{Edery:2019txq}
A.~Edery and Y.~Nakayama,
Phys. Rev. D \textbf{99}, no.12, 124018 (2019)
doi:10.1103/PhysRevD.99.124018
[arXiv:1902.07876 [hep-th]];
\bibitem{Rasanen:2018ihz}
F.~Bauer and D.~A.~Demir,
Phys. Lett. B \textbf{698}, 425-429 (2011)
doi:10.1016/j.physletb.2011.03.042
[arXiv:1012.2900 [hep-ph]];
\bibitem{Bekov:2020dww}
S.~Bekov, K.~Myrzakulov, R.~Myrzakulov and D.~S\'aez-Chill\'on G\'omez,
Symmetry \textbf{12}, no.12, 1958 (2020)
doi:10.3390/sym12121958
[arXiv:2010.12360 [gr-qc]].

\bibitem{Karam:2021sno}
A.~Karam, E.~Tomberg and H.~Veerm\"ae,
[arXiv:2102.02712 [astro-ph.CO]].
\bibitem{Gialamas:2019nly}
I.~D.~Gialamas and A.~B.~Lahanas,
Phys. Rev. D \textbf{101}, no.8, 084007 (2020)
doi:10.1103/PhysRevD.101.084007
[arXiv:1911.11513 [gr-qc]].


\bibitem{Koivisto:2005yc}
T.~Koivisto and H.~Kurki-Suonio,
Class. Quant. Grav. \textbf{23}, 2355-2369 (2006)
doi:10.1088/0264-9381/23/7/009
[arXiv:astro-ph/0509422 [astro-ph]];


\bibitem{Olmo:2015axa}
G.~J.~Olmo and D.~Rubiera-Garcia,
Universe \textbf{1}, no.2, 173-185 (2015)
doi:10.3390/universe1020173
[arXiv:1509.02430 [hep-th]].

\bibitem{Guerrero:2020uhn}
M.~Guerrero and D.~Rubiera-Garcia,
Phys. Rev. D \textbf{102}, no.2, 024005 (2020)
doi:10.1103/PhysRevD.102.024005
[arXiv:2005.08828 [gr-qc]].

\bibitem{Rosa:2020uoi}
J.~L.~Rosa, J.~P.~S.~Lemos and F.~S.~N.~Lobo,
Phys. Rev. D \textbf{101}, 044055 (2020)
doi:10.1103/PhysRevD.101.044055
[arXiv:2003.00090 [gr-qc]].

\bibitem{Odintsov:2014yaa}
S.~D.~Odintsov, G.~J.~Olmo and D.~Rubiera-Garcia,
Phys. Rev. D \textbf{90}, 044003 (2014)
doi:10.1103/PhysRevD.90.044003
[arXiv:1406.1205 [hep-th]].

\bibitem{Olmo:2019flu}
G.~J.~Olmo, D.~Rubiera-Garcia and A.~Wojnar,
Phys. Rept. \textbf{876}, 1-75 (2020)
doi:10.1016/j.physrep.2020.07.001
[arXiv:1912.05202 [gr-qc]].

\bibitem{Capozziello:2010ut}
S.~Capozziello and S.~Vignolo,
Int. J. Geom. Meth. Mod. Phys. \textbf{8}, 167-176 (2011)
doi:10.1142/S0219887811005063
[arXiv:1003.4280 [gr-qc]].

\bibitem{BeltranJimenez:2018vdo}
J.~Beltr\'an Jim\'enez, L.~Heisenberg and T.~S.~Koivisto,
JCAP \textbf{08}, 039 (2018)
doi:10.1088/1475-7516/2018/08/039
[arXiv:1803.10185 [gr-qc]].

\bibitem{Gourgoulhon:2007ue}
E.~Gourgoulhon,
[arXiv:gr-qc/0703035 [gr-qc]].

\bibitem{Lecturenotes}
M. Blau, \textit{Lecture notes on General Relativity}. http://www.blau.itp.unibe.ch/Lecturenotes.html.

\bibitem{Arnowitt:1962hi}
R.~L.~Arnowitt, S.~Deser and C.~W.~Misner,
Gen. Rel. Grav. \textbf{40}, 1997-2027 (2008)
doi:10.1007/s10714-008-0661-1
[arXiv:gr-qc/0405109 [gr-qc]].

\bibitem{Bombacigno:2019nua}
F.~Bombacigno, S.~Boudet and G.~Montani,
Nucl. Phys. B \textbf{963}, 115281 (2021)
doi:10.1016/j.nuclphysb.2020.115281
[arXiv:1911.09066 [gr-qc]].

\bibitem{Obukhov1987}
Yu~N.~Obukhov, Class. Quantum Grav. \textbf{4} 1085 (1987)

\bibitem{Gomez:2020rnq}
D.~S\'aez-Chill\'on G\'omez,
Phys. Lett. B \textbf{814}, 136103 (2021)
doi:10.1016/j.physletb.2021.136103
[arXiv:2011.11568 [gr-qc]].

\bibitem{Gionti:2020cwu}
S.~J.~Gabriele Gionti and S.~J,
Phys. Rev. D \textbf{103}, no.2, 024022 (2021)
doi:10.1103/PhysRevD.103.024022
[arXiv:2003.04304 [gr-qc]].

\bibitem{Olmo:2020fri}
G.~J.~Olmo and D.~Rubiera-Garcia,
Class. Quant. Grav. \textbf{37}, no.21, 215002 (2020)
doi:10.1088/1361-6382/abb924
[arXiv:2007.04065 [gr-qc]].

\bibitem{Guerrero:2021pxt}
M.~Guerrero, G.~J.~Olmo and D.~Rubiera-Garcia,
[arXiv:2102.00840 [gr-qc]].

\bibitem{Madsen:1989rz}
M.~S.~Madsen and J.~D.~Barrow,
Nucl. Phys. B \textbf{323}, 242-252 (1989)
doi:10.1016/0550-3213(89)90596-8

\bibitem{Nojiri:2000kh}
S.~Nojiri, S.~D.~Odintsov and S.~Ogushi,
Phys. Rev. D \textbf{62}, 124002 (2000)
doi:10.1103/PhysRevD.62.124002
[arXiv:hep-th/0001122 [hep-th]].

\bibitem{Dyer:2008hb}
E.~Dyer and K.~Hinterbichler,
Phys. Rev. D \textbf{79}, 024028 (2009)
doi:10.1103/PhysRevD.79.024028
[arXiv:0809.4033 [gr-qc]].

\bibitem{Guarnizo:2010xr}
A.~Guarnizo, L.~Castaneda and J.~M.~Tejeiro,
Gen. Rel. Grav. \textbf{42}, 2713-2728 (2010)
doi:10.1007/s10714-010-1012-6
[arXiv:1002.0617 [gr-qc]].

\bibitem{Senovilla:2013vra}
J.~M.~M.~Senovilla,
Phys. Rev. D \textbf{88}, 064015 (2013)
doi:10.1103/PhysRevD.88.064015
[arXiv:1303.1408 [gr-qc]].

\bibitem{Chu:2021uec}
C.~S.~Chu and H.~S.~Tan,
[arXiv:2103.06314 [hep-th]].

\bibitem{Deruelle:2007pt}
N.~Deruelle, M.~Sasaki and Y.~Sendouda,
Prog. Theor. Phys. \textbf{119}, 237-251 (2008)
doi:10.1143/PTP.119.237
[arXiv:0711.1150 [gr-qc]].

\bibitem{delaCruz-Dombriz:2014zaa}
\'A.~de la Cruz-Dombriz, P.~K.~S.~Dunsby and D.~Saez-Gomez,
JCAP \textbf{12}, 048 (2014)
doi:10.1088/1475-7516/2014/12/048
[arXiv:1406.2334 [gr-qc]].

\bibitem{Senovilla:2014kua}
J.~M.~M.~Senovilla,
Class. Quant. Grav. \textbf{31}, 072002 (2014)
doi:10.1088/0264-9381/31/7/072002
[arXiv:1402.1139 [gr-qc]].

\bibitem{Reina:2015gxa}
B.~Reina, J.~M.~M.~Senovilla and R.~Vera,
Class. Quant. Grav. \textbf{33}, no.10, 105008 (2016)
doi:10.1088/0264-9381/33/10/105008
[arXiv:1510.05515 [gr-qc]].


\end{thebibliography}
\end{document}